\documentclass[aps,prd,twocolumn,nofootinbib]{revtex4}

\usepackage{graphicx}
\usepackage{amsmath,amssymb,latexsym}
\usepackage{bm}
\usepackage{epsfig}
\usepackage{dcolumn}

\newcommand{\be}{\begin{equation}}
\newcommand{\ee}{\end{equation}}

\def\4he{$^4$He}
\def\3he{$^3$He}
\def\7li{$^7$Li}

\newcommand\la{\lower0.6ex\vbox{\hbox{\ensuremath{\buildrel{\textstyle<}\over{\sim}\ }}}}
\newcommand\ga{\lower0.6ex\vbox{\hbox{\ensuremath{\buildrel{\textstyle>}\over{\sim}\ }}}}

\def\lsim{\mathrel{\raise.3ex\hbox{$<$\kern-.75em\lower1ex\hbox{$\sim$}}}}
\def\gsim{\mathrel{\raise.3ex\hbox{$>$\kern-.75em\lower1ex\hbox{$\sim$}}}}

\begin{document}
\DeclareGraphicsExtensions{.eps,.ps}

\title{Upward shower rates at neutrino telescopes directly determine the 
neutrino flux}
\author{S~Hussain$^{1}$, D.~Marfatia$^{2}$ and D.~W.~McKay$^{2}$}
\affiliation{$^1$Department of Physics and Astronomy, University of Delaware, Newark, DE 19716}
\affiliation{$^2$Department of Physics and Astronomy, University of Kansas, Lawrence, KS 66045}

\begin{abstract}

We show that the rate for upward showers from an isotropic cosmic neutrino flux at neutrino 
telescopes like IceCube is independent of the neutrino-nucleon cross section. 
For bins that span a relatively narrow range in energy, neither scaling the cross section, nor 
changing its power-law energy behavior affects the upward shower rate, which depends only on the flux. 
The neutrino flux can be completely known since its spectral shape can be determined by comparing the rates 
in neighboring bins. We also show that the 
downward shower rate varies linearly 
with cross section with a proportionality constant determined by the energy-dependence of the cross section, independent of the power-law behavior of the flux.  The 
normalization and energy dependence of the cross section can be found by comparing the downward rates in neighboring bins.


\end{abstract}

\maketitle

{\bf{Introduction.}}
Neutrino telescopes looking for high energy cosmic neutrinos are already ruling out~\cite{newrice}
more aggressive flux models from diffuse sources~\cite{agn} and of cosmogenic origin~\cite{gzk}. 
The limits placed are within the framework of the Standard Model (SM). 
However, it is conceivable that new physics manifests itself in neutrino-nucleon interactions
at high energies, enhancing the interest in and difficulty of studying high energy neutrinos.
The task of simultaneously determining the cosmic flux and the high energy cross section is a frustrating task 
since laboratory sources to examine
neutrino interactions in controlled experiments at the needed energies are
out of the question. Astrophysical fluxes cannot be tweaked in the control room, and the 
highest energies are beyond the reach of even the Large Hadron Collider. The challenge to find 
observables that lead to efficient extraction of the fluxes and interactions is the motivation for this paper, 
as it has been for a number already in print~\cite{kw,dan1,hmms,Borriello:2007js}.  

We investigate the effect on shower rates 
of varying the power-law energy dependence of the cosmic neutrino flux and of the cross section 
in addition to the overall strength of the cross section. We illustrate how to extract
complete knowledge (normalization and energy dependence) of the cosmic flux and the neutrino-nucleon
cross section.

{\bf Methodology.} Using reasoning similar to that of Ref.~\cite{hmms}, 
the rate for showers in a volume detector centered at depth $d$ can be written as
\begin{equation}
\Gamma =\frac{d\phi}{d\Omega}\frac{\pi A_p}{(R-d)\lambda_d}
\int_{a}^{b} \lambda e^{-\frac{l}{\lambda}}
\big(e^{\frac{s}{\lambda}}-1\big)\bigg({L_{h}^2 \over l^2}+1\bigg) dl\,,
\label{gamma}
\end{equation}
where $\frac{d\phi}{d\Omega}$ is the differential flux, $A_p$ is the area projected against 
the neutrino direction, $R$ is the radius of the earth, 
$s$ is the scale of the detector size, $l$ is the chord length traversed by a neutrino, 
and $L_h \equiv \sqrt{d(2R-d)}$ is the horizontal distance from the detector to the ``horizon''. 
Since the nucleon density $n$ varies along a chord, we use an attentuation length $\lambda=1/(n\sigma_t)$ 
that is averaged over $l$; $\sigma_t$ is the total interaction cross section. To emphasize that 
the interaction length at the detector is not averaged over a chord segment, we denote
 it as $\lambda_d$.  
 For up (down) going events, the integration limits are $a=L_{h}$, $b=2R-d$ ($a=d$, $b=L_h$).  The nadir angle that divides upward from downward events is therefore $90^\circ$, as measured from the detector at depth $d$.  We find that changing the definition of upward versus downward events by $\pm 3^\circ$ has no effect on our conclusion that the upward event rate is determined by the flux alone.

While Eq.~(\ref{gamma}) accounts for the depth of the detector and the Earth's density profile ~\cite{prem}, it does
not include downscattering or neutrino regeneration effects, which however, are typically of order 10\% and 
irrelevant for our considerations. 
The full propagation requires the solution to coupled integro-differential equations, and, as shown 
in Ref.~\cite{hmms}, the numerical calculations support the conclusions one draws from the analytic result.

We employ the description of physics models introduced
in Ref.~\cite{hmms}. 
The strength of charged current (c) and neutral current (n) interactions are parameterized 
by $\alpha_c=\sigma_c/\sigma_t^{\rm SM}$ and $\alpha_n=\sigma_n/\sigma_t^{\rm SM}$, respectively, and are assumed to
be energy-independent within a bin. The SM cross
sections correspond to $(\alpha_c,\alpha_n)=(r_c,r_n)$, where $r_i=\sigma_i^{\rm SM}/\sigma_t^{\rm SM}$. 
We consider new physics that scales the SM charged current and neutral current cross sections by the same amount,
$(\alpha_c,\alpha_n)=(\alpha r_c,\alpha r_n)$, with inelasticity the same as in the SM. We allow $\alpha$ to take
values from 0.2 to 10. We also allow for changes in the power law behavior of the SM, which is approximately 
$\sigma_{SM}(E_\nu) \sim E_\nu^{0.36}$~\cite{raj}. 
We adopt a simple form $\sigma(E_\nu) \sim E_\nu^{0.36+\beta}$ with
$\beta$ ranging from $-0.36$ to $1$, and normalize so that $\beta=0$ reproduces the rate for $\alpha=1$.\footnote{Note that power law growths of $E^{1 - 1.5}$ are typical of inelastic, hadronic shower dominated cross sections from large extra dimension physics in the energy bins 
under study~\cite{fsag}.}  

We consider neutrino fluxes that fall with energy as 
$E_\nu^{-\gamma}$, with $\gamma=1,2,3$ and normalize the event rates at $\alpha=1$ to that for an 
isotropic total flux of neutrinos and antineutrinos $\frac{d\phi}{d\Omega} = 6 \cdot 10^{-8} 
(E_\nu/{\rm GeV})^{-2}$~(cm$^2$.s.sr.GeV)$^{-1}$~\cite{wb} 
with an assumed flavor ratio at Earth of 1:1:1; the Waxman-Bahcall (WB) flux. 
We integrate rates over bins
of width $\Delta\log_{10}(E_\nu/$GeV$)=0.5$. 
 
\begin{figure}
\includegraphics[angle=-90,width=3.15in]{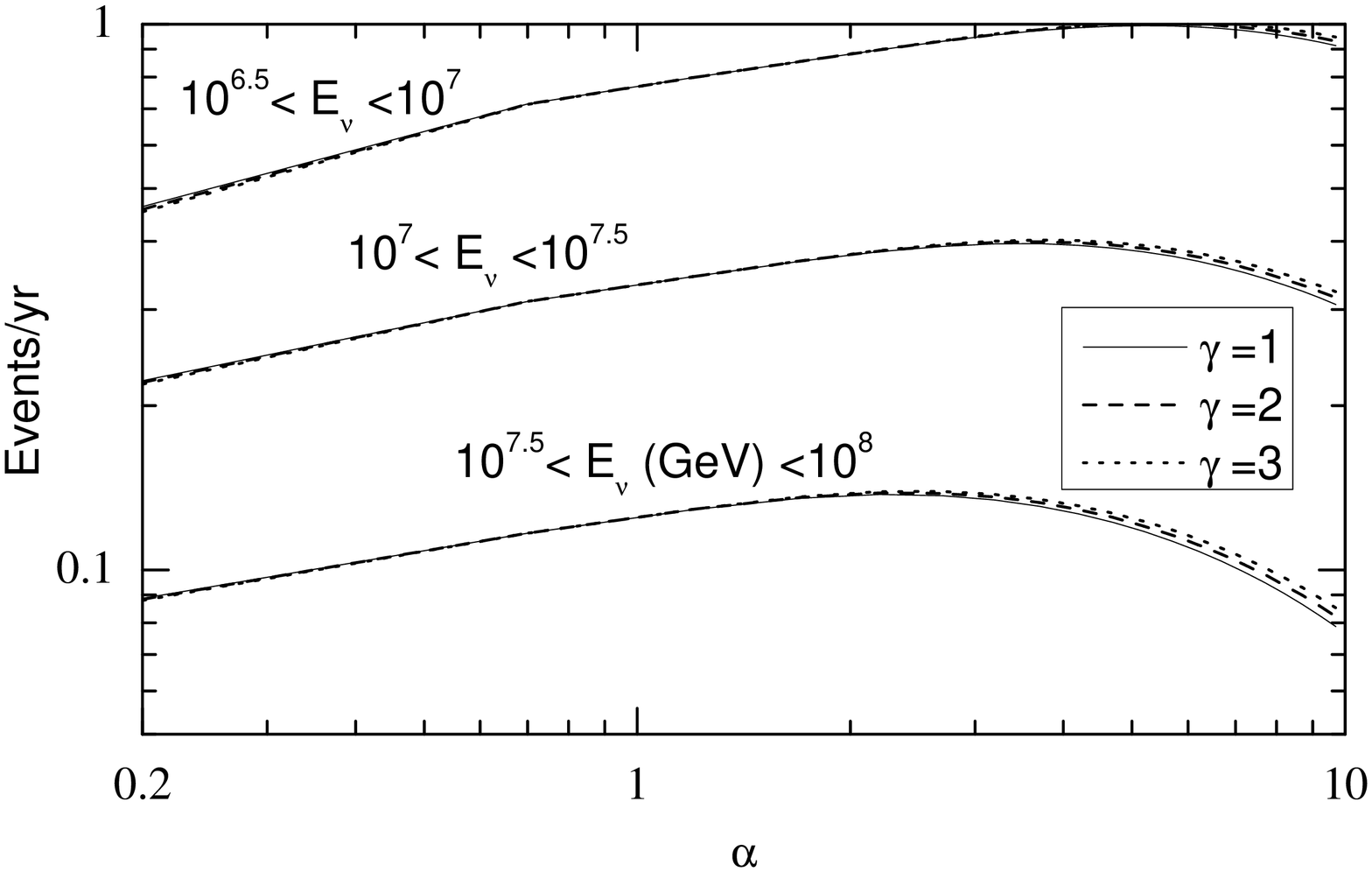}
\caption{ Event rates (in three representative energy bins) vs. $\alpha$, 
for fluxes falling with energy as $E_\nu^{-\gamma}$.
The plots are normalized at $\alpha=1$ to the WB event rate.  
}
\label{fig:fig1}
\end{figure}

{\bf{Dependence of shower rates on $\alpha$ and $\gamma$.}}
We first set $\beta=0$.
In Fig.~\ref{fig:fig1}, we plot the rate for upward showers in energy bins spaced equally in logarithm,
 $10^{6.5}<E_{\nu}<10^7$~GeV, 
$10^7 < E_{\nu} < 10^{7.5}$~GeV and $10^{7.5}<E_{\nu}<10^8$~GeV, as a function of $\alpha$ for $\gamma=1,2$ and $3$. For convenience in comparing the shape of the functional dependence of rates on $\alpha$ for different input fluxes, we have normalized the $\gamma=1$ and $3$ curves to the WB value at $\alpha$ = 1, in each energy bin.
The resulting shapes are nearly identical, independent of $\gamma$. 
For all three fluxes, the variation of the rate is a factor 2 or less over the factor 50 change in cross section strength. The variation is less than 50\% for
$\alpha$ between 0.5 and 5.  The flux determined by using the 
SM cross section in the analysis is the correct one, since any other cross section with $\beta=0$ would give 
the same value.


\begin{figure}
\includegraphics[angle=-90,width=3.15in]{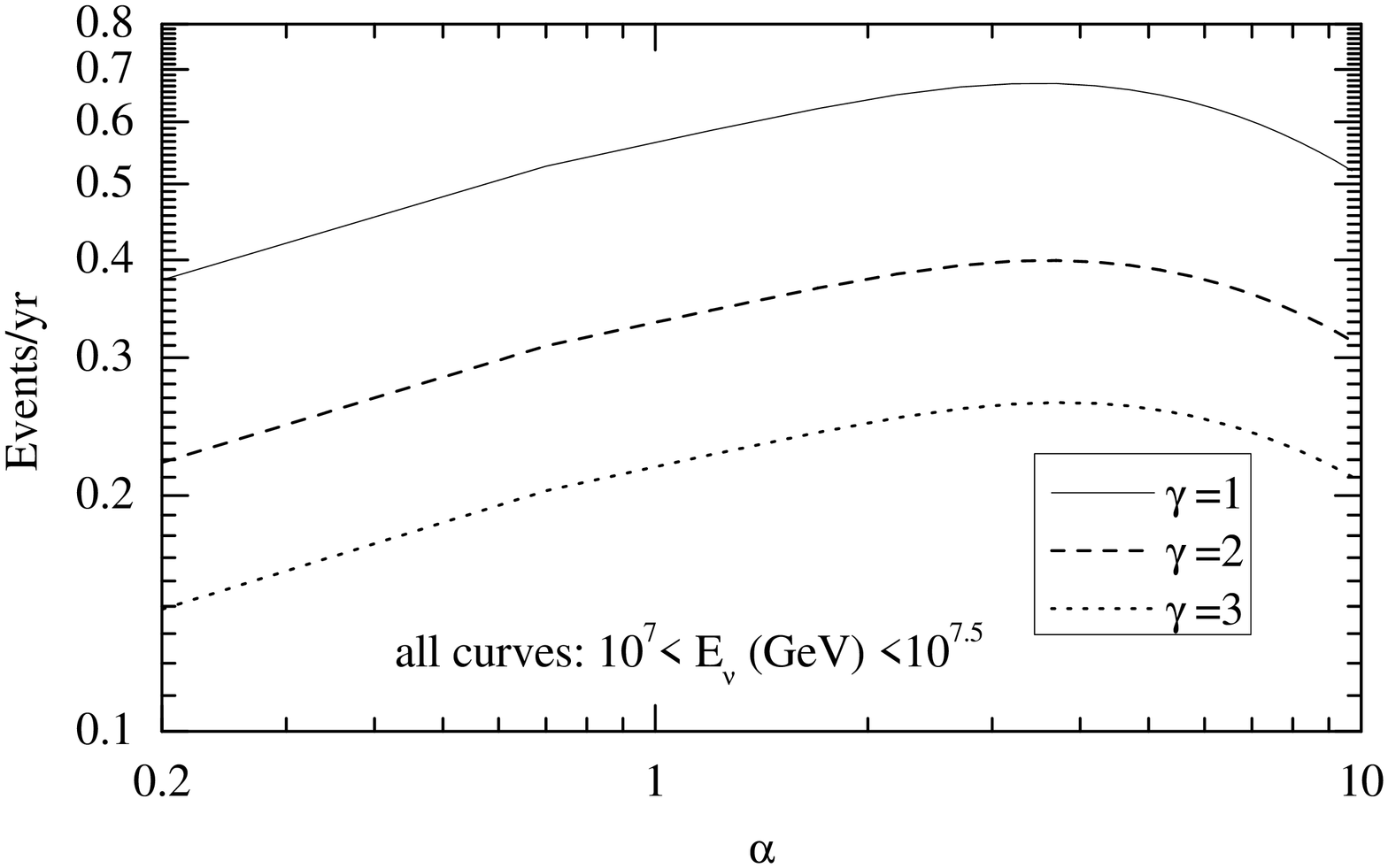}
\caption{ Event rates vs. $\alpha$, for fluxes falling with energy as $E_\nu^{-\gamma}$, 
with their true normalization for the middle energy bin, $10^7 < E_\nu < 10^{7.5}$~GeV.
}  
\label{fig:fig2}
\end{figure}

In Fig. ~\ref{fig:fig2} we show the curves for the middle bin with their true normalization, since the curves in each energy bin in Fig.~\ref{fig:fig1}  are normalized to the value appropriate to the WB flux with the SM cross section for convenience in comparing the shapes.  Putting the information from the two figures together, we see clearly that \emph{the spectral index of the flux affects the rate in a bin, but it does not affect the independence of the rate from $\alpha$}.

\begin{figure}
\includegraphics[angle=-90,width=3.15in]{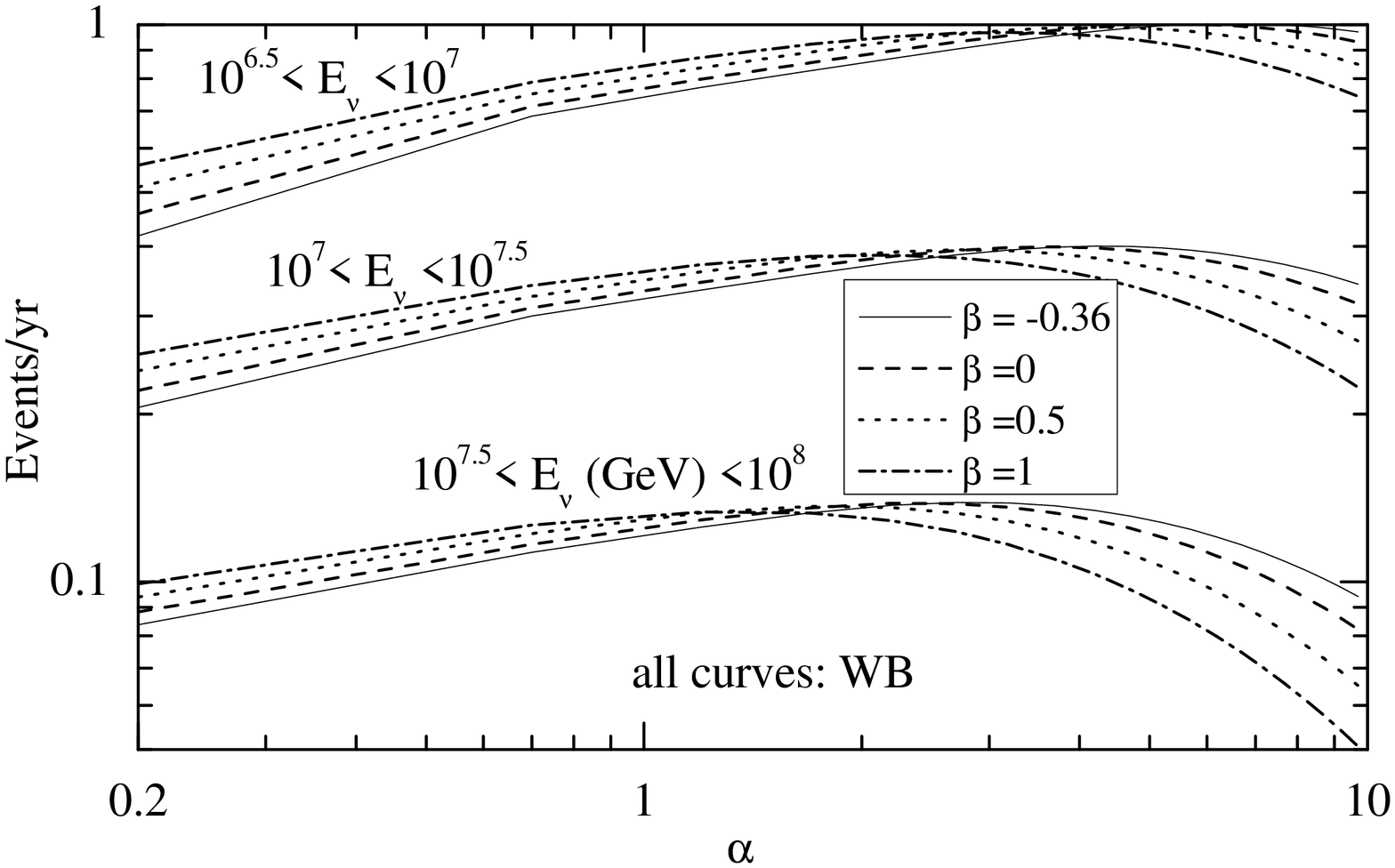}
\caption{Upward shower event rates vs. $\alpha$ for the WB flux for four values of $\beta$.
The curves are normalized such that the SM rate is reproduced for $\beta=0$ and $\alpha=1$.}
\label{fig:fig4}
\end{figure}

{\bf{Dependence of shower rates on $\alpha$, $\beta$ and $\gamma$.}}
We now explore the effect of varying $\alpha$, $\beta$ and $\gamma$ simultaneously. 
In Fig.~\ref{fig:fig4} we show how the upward event rates in our three representative bins 
are affected for $\beta = -0.36, 0, 
0.5$ and $1.0$. We only present the results for the WB flux ($\gamma=2$) because for a given 
$\beta$, choosing $\gamma=1,3$ makes a barely perceptible difference. Allowing $\beta$ to range from $-0.36$ to $1$ 
produces a shift in shape of the curves, but the values stay within the same 50\% range around the SM value 
for any value of $\alpha$. Neither energy dependence nor normalization of the cross section appreciably modifies 
the up shower event rate in our ``fiducial'' bins.  Changing the energy dependence of the flux also 
makes little difference.  {\it The upward shower rate per bin is determined by the flux alone.}  
The energy dependence of the flux can be determined by comparing rates in neighboring energy bins, as
we demonstrate later.

\begin{figure}
\includegraphics[angle=-90,width=3.15in]{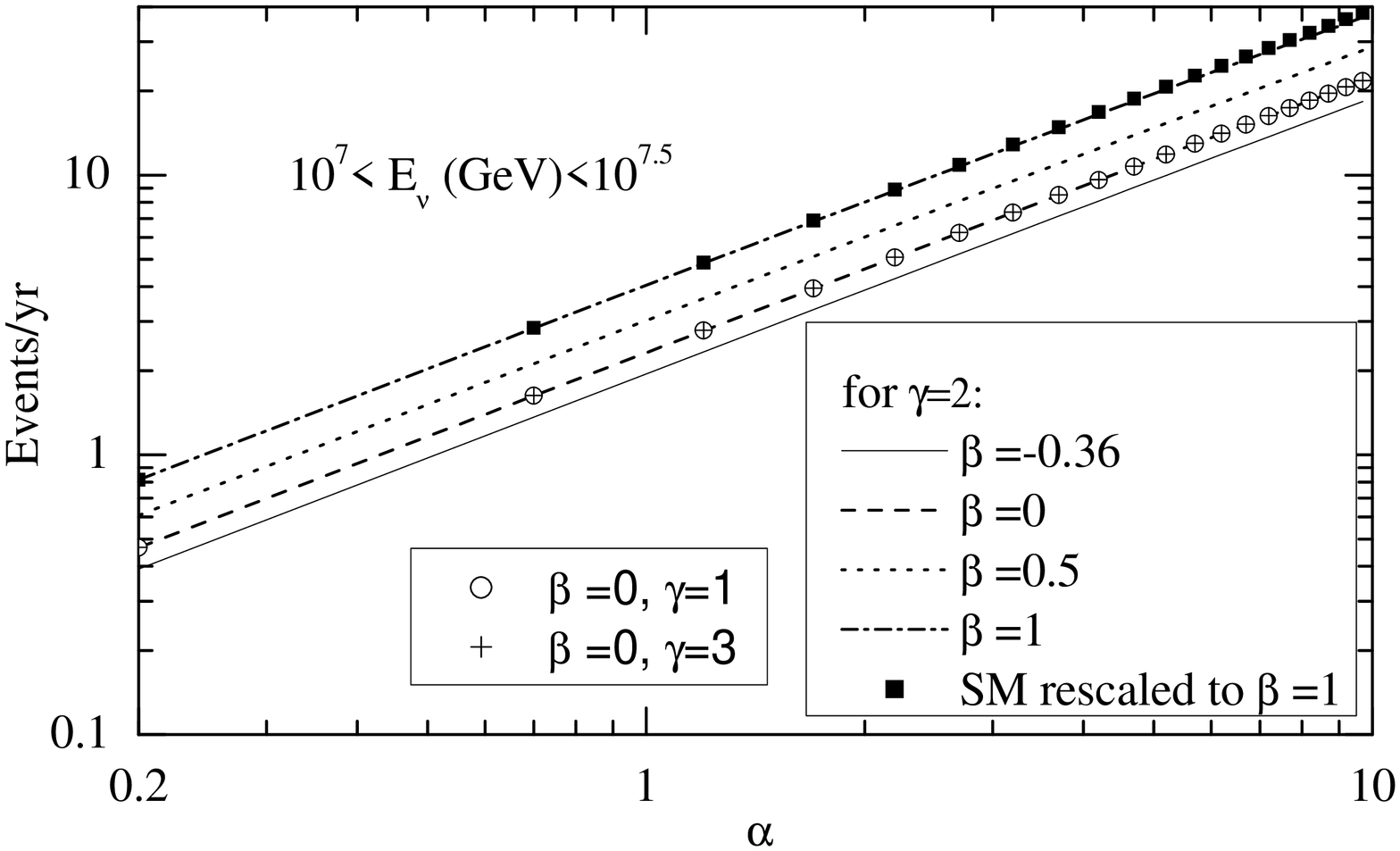}
\caption{Similar to Fig.~\ref{fig:fig4}, but for downward showers only 
in the energy bin $10^7 < E_\nu < 10^{7.5}$~GeV. 
The squares almost on 
top of the dot-dashed line is the SM rate ($\beta = 0$) rescaled to agree with the $\beta=1$ rate at $\alpha= 0.2$. 
The slight slump of the $\beta=1$ curve relative to the SM indicates that 
absorption at the highest cross section values is starting to appear. 
Note that the event rate for $\beta=0$ is unaffected by changing $\gamma$.}
\label{fig:fig5}
\end{figure}

The down shower rates as a function of $\alpha$ for different $\beta$ and $\gamma$ are shown in Fig.~\ref{fig:fig5}.  
The expected proportionality to $\alpha$~\cite{hmms} is clear, with values of the rates determined by $\beta$.
We see that the event rate for $\beta=0$ is unaffected by 
changing $\gamma$, since the dashed line, the open circles and the plus signs lie on top of each other. 


{\bf{Flux and cross section from data.}}
Given the energy dependence of the flux,  the measured rate of upward showers in an energy bin
 will determine the normalization of the flux.  It is the only other quantity, apart from the known acceptance and earth density profile,  
that the rate in a bin depends upon over a wide range of model assumptions.  
To obtain the energy dependence of the flux, one compares the rates in neighboring bins to extract $\gamma$ 
in that range of energy.  To the extent that the neighboring bins which span the energy ranges $E_{i-1}-E_i$ and 
$E_i-E_{i+1}$, are approximated by the same 
power law in energy, and the acceptance, $A$, is roughly constant in each bin, the ratio of upward shower rates
 in neighboring bins is for $\gamma \neq 1$,
\begin{equation}
{\Gamma(i-1,i) \over \Gamma(i,i+1)} = \frac{E_{i}^{\gamma-1}-E_{i-1}^{\gamma-1}}{E_{i+1}^{\gamma-1}-E_{i}^{\gamma-1}} 
\bigg(\frac{E_{i+1}}{E_{i-1}}\bigg)^{\gamma-1}\frac{A_{i-1,i}}{A_{i,i+1}}\frac{\rho_{i,i+1}}{\rho_{i-1,i}}\,,
\end{equation} 
where the rate in a bin is denoted by $\Gamma(i,i+1)$, the acceptance by $A_{i,i+1}$ and the 
average density ratio $\rho_{Earth}/\rho_{ice}$ met by the incoming flux contributing to events in a bin by 
$\rho_{i,i+1}$. For the special case $\gamma=1$, the explicitly energy dependent factors are replaced by $\ln(E_{i}/E_{i-1})/\ln(E_{i+1}/E_i)$ (When the bins are equally spaced, the ratios will all be 1.) 
In the table, 
we show the result of this simple analysis 
for simulated up shower data~\cite{hmms} with the IceCube effective volume~\cite{luis&francis} and the WB flux.
\begin{table}
\begin{tabular}{|l|c|}  \hline
bin[$E_{i-1}$, $E_{i}$]/bin[$E_{i}$, $E_{i+1}$]  & $\Gamma[i-1, i]/\Gamma[i,i+1]$ \\ \hline
bin[$10^{6.5}$, $10^7$]/bin[$10^7$, $10^{7.5}$ ]      & 2.7   \\ \hline
bin[$10^7$, $10^{7.5}$]/bin[$10^{7.5}$, $10^8$ ]      & 2.8   \\ \hline
bin[$10^{7.5}$, $10^8$]/bin[$10^8$, $10^{8.5}$ ]      & 2.9   \\ \hline
bin[$10^8$, $10^{8.5}$]/bin[$10^{8.5}$, $10^9$ ]      & 2.9   \\ \hline
\end{tabular}
\caption{Ratio of upward shower rates in adjacent bins for the WB flux from a simulation of the IceCube detector.}
\end{table}
We list the ratio of rates in neighboring bins from 
simulated data of upward showers in an ``IceCube - like'' volume detector. The actual values are taken for the 
SM cross section, but changing the cross section changes the ratios very little, as we have shown.  
Equation 2 implies that for all bins, the values for the $\gamma=1,2$ and $3$ flux spectra are $10^0$, $10^{0.5}$ and $10^1$, respectively, from the explicit energy factor, with reductions of order $10\%$ coming from the acceptance and density factors.  
Clearly the values are consistent only for $\gamma=2$, which is the spectrum used to generate the data. 

\begin{figure}
\includegraphics[angle=-90,width=3.15in]{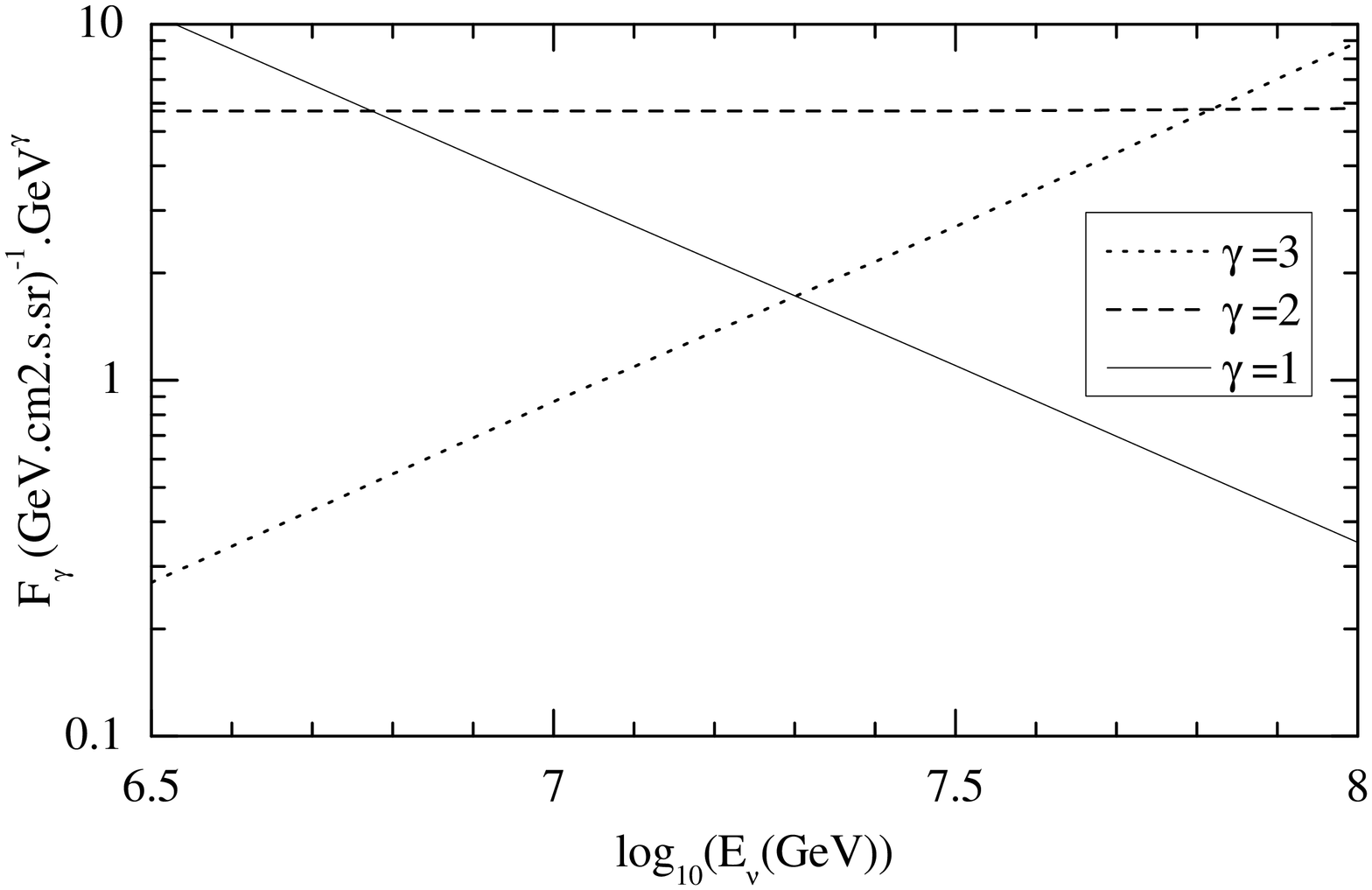}
\caption{ Flux coefficient, $F_\gamma$, for fluxes falling with energy as $E_\nu^{-\gamma}$. 
To display on the same plot, we multiplied the coefficients by $10^{15}$, $10^8$, and 1. The event rate data used in 
the analysis is generated from the WB flux model, which has $\gamma = 2$.  
Five energy bins from $10^{6.5}$ to $10^{9}$~GeV of width $\Delta\log_{10}(E_\nu/$GeV$)=0.5$ are chosen 
for illustration.  
The coefficient must be the same in each bin to be consistent with data.  Clearly, the WB case is reproduced while the $\gamma=1$ and 3 cases are strongly excluded.}
\label{fig:fig6}
\end{figure}

A graphical representation is shown in Fig.~\ref{fig:fig6}. We use the up shower event rates generated by the SM cross section and WB flux in our representative bins to extract the overall coefficients, $F_{\gamma}$, of the 
$F_{\gamma} E_\nu^{-\gamma}$ flux spectra for $\gamma=1$, 2 and 3. Factors appropriate to their different dimensions were applied to display them all on the same graph.  With these simple input models, the data choose the one whose coefficient is the same in each bin.  As shown in the figure, the $\gamma=1$ and 3 cases are easily excluded.  As indicated by our earlier analysis, this result will be the same even for widely different cross sections.

The flux in a given bin can be estimated from Eq.~(\ref{gamma}), given the value of 
the effective volume $V_{eff}= s\times A_{p}$, $\Gamma_{up}$ for the bin, the Earth's radius and the ratio of
the average density of earth to density of target (ice). The average density the incoming neutrinos ``see'' depends mildly on energy.  In the bin between $10^7$~GeV and $10^{7.5}$~GeV, 
simulation gives $\sim 0.3$  events/year for cross sections with $\alpha$ between 0.2 and 10, 
with the normalization of Ref.~\cite{wb}.  We use $V_{eff}\sim 2$ km$^3$~\cite{luis&francis}  
and average density ratio $\rho_{Earth}/\rho_{ice} \sim 2-3$ for the 
crust/mantel segment that dominates the upcoming flux, and find the flux normalization to be 
$\sim 6-9  \cdot 10^{-15}$ (cm$^2$$\cdot$ s $\cdot$ sr)$^{-1}$, 
which agrees roughly with the input normalization.  

\begin{figure}
\includegraphics[angle=-90,width=3.15in]{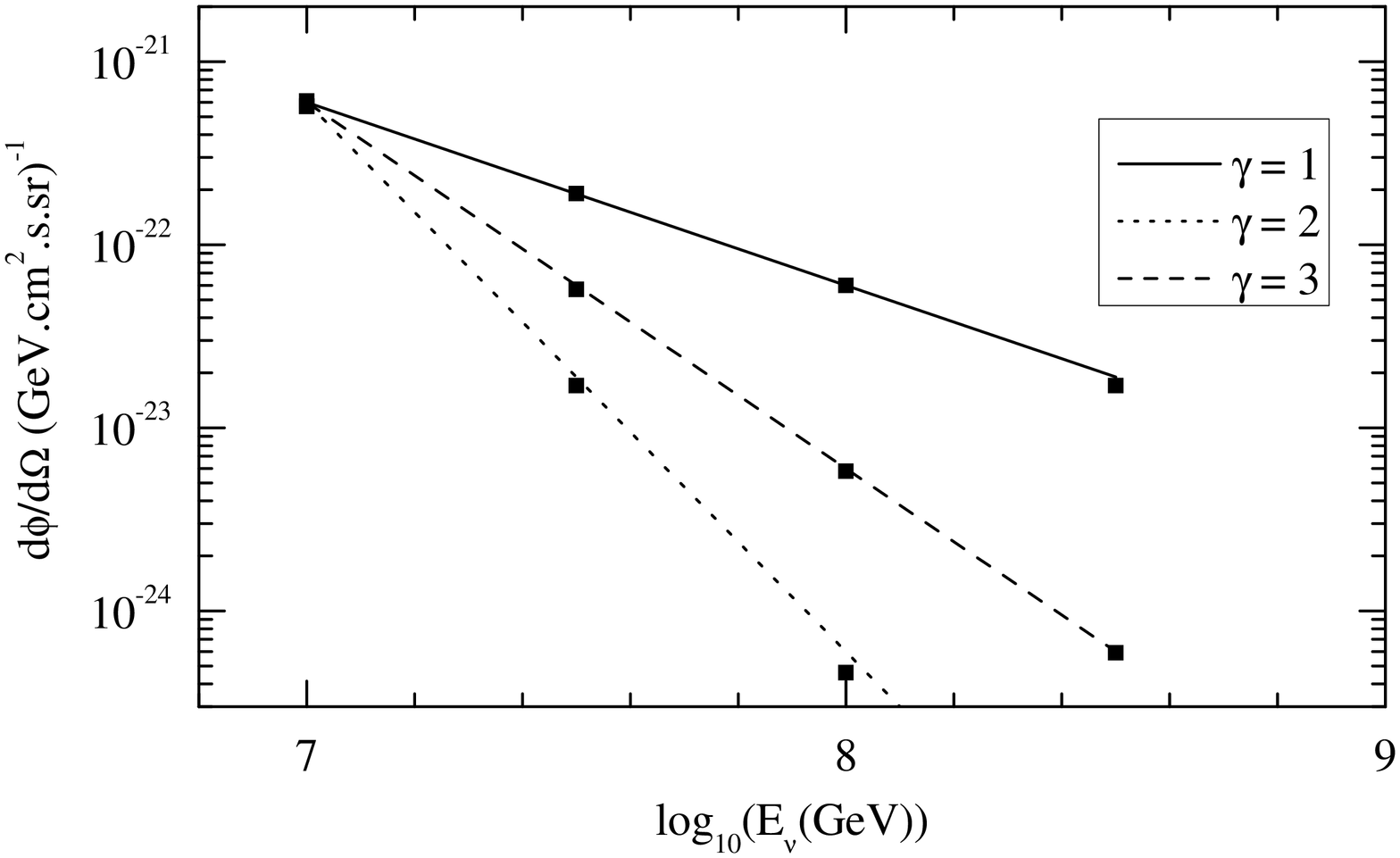}
\caption{ The squares show the output values of the differential flux extracted 
 by the procedure of Eq.~(\ref{extract}) (except 
for $\gamma \simeq 1$), 
from the numbers of up shower events calculated in our model.   
The event rate data that are used in the analysis are generated from flux models 
with $\gamma = 1$, 2 and 3, all normalized to the WB flux at $10^7$ GeV, and the cross
sections have $\beta=0$ and  $\alpha = 1$, 1 and
6, respectively.   
Five energy bins from $10^{6.5}$ to $10^{9}$~GeV of width $\Delta\log_{10}(E_\nu/$GeV$)=0.5$ 
are chosen for the analysis, 
which then produces flux values at four energies.}  
\label{fig:fig7}
\end{figure}

{\bf Extracting the differential flux from data.} 
Expanding on this question of flux measurement, we next estimate the differential flux index  \emph{directly from data} by the following prescription.  Given data for upward shower rates in several adjoining energy bins, we refer again to Eq.~(\ref{gamma}) to motivate the definition, for $\gamma \neq$ 1, 

\begin{equation}
\frac{d\phi(E_i)}{d\Omega} = \frac{R}{\pi}\frac{\Gamma(i-1,i)\frac{\rho_{i-1,i}}{V_{eff}(i-1/2)}-\Gamma(i,i+1)\frac{\rho_{i,i+1}}{V_{eff}(i+1/2)}}{E_{i+1/2}-E_{i-1/2}}\,,
\label{extract}
\end{equation} 
 where the $i \pm 1/2$ notation indicates that the quantity is 
evaluated at the midpoint (in logarithm) of the relevant bin. 
The energy assigned to the flux is between two adjoining bins for 
plotting purposes. The flux index is found by fitting the points
obtained from Eq.~(\ref{extract}). The normalization can then be obtained from 
the event rate in a particular bin. If  
the observed rates are nearly the same in each bin, $\gamma \simeq 1$  applies and
Eq.~(\ref{extract}) is irrelevant. 
A fit near $\gamma = 1$ can be found and 
the normalization determined from any bin in this range. 
Applying this prescription to our analytic model for three fluxes, 
we obtain the values for the differential fluxes shown by the squares in Fig.~\ref{fig:fig7}.  
The lines are the input fluxes normalized to the WB flux at $10^7$ GeV. 
The procedure of Eq.~(\ref{extract}) is obviously successful 
in extracting the original flux. We have checked that the procedure is effective for other 
flux and cross section combinations, 
as expected from the results shown in Figs.~\ref{fig:fig1},~\ref{fig:fig2} and~\ref{fig:fig4}. 

The ratio of downward events to upward events in a given energy bin is potentially a clean way to extract information 
about the total cross section in that bin~\cite{dan1,doug}.  
Using Eq.~(\ref{gamma}) in the usual energy range,  
with average density $2-3 \cdot N_{A}$, and a ratio of down to up events $\sim 7$, we find
$\sigma = 3-5 \cdot 10^{-33}$ cm$^2\times \alpha$.  This agrees nicely with the SM 
value at $\alpha = 1$~\cite{raj} in this energy interval.

{\bf{Summary.}}
Our study confirms the linear dependence of downward events on the product of 
cross section and flux values and establishes the robustness of the linear dependence of upward shower event rates on flux alone.  Comparing rates in neighboring bins yields the energy and normalization of flux and cross section.
We have focused on the systematics of the analysis by using a model which captures the main features of the upcoming shower events.  It has the essential elements of a full simulation, but of course does not pretend to provide the detail necessary for analysis of real data. We have not addressed the tough question of reconstructing neutrino energy, the variable we adopt directly here, from the energy measured in a shower. 

Our message here is that distinguishing upcoming showers from downgoing ones, with reasonable energy resolution can have a big payoff in scientific discovery. If the charged current events are proportional to the total cross section, the discussion we presented applies equally well to them.   As shown here, the \emph{upcoming showers provide a remarkably clean way to isolate the flux independently of cross section,  over a wide range of energies}.


{{\bf Acknowledgments.}}
We thank D.~Seckel for participating in the early stages of this work.
This research was supported 
by DOE Grant No.~DE-FG02-04ER41308, 
by NASA Grant No.~NAG5-5390, and by NSF Grant Nos.~PHY-0544278 and 
OPP-0338219.

\end{document}